\documentclass[10pt,twocolumn,twoside]{IEEEtran} 
\usepackage{ifpdf}

\usepackage{cite}

\ifCLASSINFOpdf
   \usepackage[pdftex]{graphicx}
   \graphicspath{{./figs/pdf/}{./figs/jpeg/}}
   \DeclareGraphicsExtensions{.pdf,.jpeg,.png}
\else
   \usepackage[dvips]{graphicx}
   \graphicspath{{./figs/eps/}}
   \DeclareGraphicsExtensions{.eps}
\fi
\usepackage[cmex10]{amsmath}
\usepackage{array}

\ifCLASSOPTIONcompsoc
  \usepackage[caption=false,font=normalsize,labelfont=sf,textfont=sf]{subfig}
\else
  \usepackage[caption=false,font=footnotesize]{subfig}
\fi
\usepackage{fixltx2e}
\usepackage{longtable}          
\usepackage{multirow}
\usepackage{framed}
\usepackage[table]{xcolor}
\usepackage{epstopdf}
\usepackage{amssymb}
\usepackage{enumerate}
\usepackage{dsfont}
\usepackage{amsthm} 

\definecolor{CorCol}{rgb}{0,0,0}

\newtheorem{corol}{Corollary}
\newtheorem{Def}{Definition}
\newtheorem{Prop}{Proposition}
\newtheorem{lem}{Lemma}

\newcommand{\tran}{^{\mathsf{T}}}
\newcommand{\expectx}{\mathds{E}_{\mathbf{x}_k}\!}
\newcommand{\expectz}{\mathds{E}_{\mathbf{z}_{1:k}}\!}
\newcommand{\xmmse}{\hat{\mathbf{{x}}}_{k}^{\mathrm{MMSE}}}
\newcommand{\Pmmse}{P_{k|k}^{\mathrm{MMSE}}}
\newcommand{\corcol}[1]{\textcolor{CorCol}{#1}}
\begin{document}
\allowdisplaybreaks

%
\title{Analytic MMSE Bounds in Linear Dynamic Systems with Gaussian Mixture Noise Statistics}
%
%
%

\author{Leila~Pishdad,
        Fabrice~Labeau
\thanks{The authors are with the Department of Electrical and Computer Engineering, McGill University, Montreal, QC H3A 0E9, Canada (e-mail: leila.pishdad@mail.mcgill.ca; fabrice.labeau@mcgill.ca)}
}

\maketitle

\begin{abstract}
Using state-space representation, mobile object positioning problems can be described as dynamic systems, with the state representing the unknown location and the observations being the information gathered from the location sensors. For linear dynamic systems with Gaussian noise, \corcol{the} Kalman filter provides the Minimum Mean-Square Error (MMSE) state estimation by tracking the posterior. Hence, by approximating non-Gaussian noise distributions with Gaussian Mixtures (GM), a bank of Kalman filters or Gaussian Sum Filter (GSF), can provide the MMSE state estimation. However, the MMSE itself is not analytically tractable. Moreover, the general analytic bounds proposed in the literature are not tractable for GM noise statistics. Hence, in this work, we evaluate the MMSE of linear dynamic systems with GM noise statistics and propose its analytic lower and upper bounds. We provide two analytic upper bounds which are the Mean-Square Errors (MSE) of implementable filters, and we show that based on the shape of the GM noise distributions, the tighter upper bound can be selected. We also show that for highly multimodal GM noise distributions, the bounds and the MMSE converge. Simulation results support the validity of the proposed bounds and their behavior in limits. 
\end{abstract}
%
%
%
%
%
%
%
%
%
%
\section{Introduction}
Due to the growing use of location-aware applications and ubiquitous systems, there has been an increasing need for more precise and accurate positioning systems. For outdoor scenarios, GNSS technologies can provide approximations with errors of 1 to 5 meters~\cite{hightower_location_2001}, which is suitable for outdoor applications. However, due to the increased multipath fading and non-line-of-\corcol{sight} (NLOS) conditions of indoor environments, indoor positioning and location tracking can be more challenging~\cite{pahlavan_indoor_2002}. Therefore, Bayesian tracking techniques have been widely used to improve the location estimation by tracking the \textit{posterior}, or \textit{belief function}~\cite{fox_bayesian_2003}. Specifically by using state-space representation for positioning problems, the motion and sensor models are defined by the process and measurement equations, respectively. Using these and the available sensor measurements, Bayesian tracking techniques, can approximate the posterior, hence provide the MMSE state estimator, as it is the expected value of the posterior~\cite{poor_introduction_1994,bar2001estimation}. For linear systems with Gaussian process and measurement noise, \corcol{the} Kalman filter can optimally track the posterior~\cite{ho_bayesian_1964,arulampalam_tutorial_2002}, and provide the MMSE state estimator~\cite{anderson1979optimal}. However, in many scenarios, neither the process, nor the measurement noise are Gaussian. For instance, in~\cite{pishdad_optimal_2013}, a non-Gaussian trimodal process noise distribution is used, corresponding to the three different states of motion: constant velocity, accelerating, and decelerating. Additionally, mutipath fading effects could result in multimodal non-Gaussian measurement noise distributions~\cite{pishdad_indoor_2012,pishdad_optimal_2013,kung-chung_lee_comparison_2010}.

Since any distribution can be approximated by a Gaussian Mixture (GM) as closely as desired~\cite{anderson1979optimal} and the estimation will be asymptotically unbiased~\cite{1962KernelDensityEstimation}, Gaussian sum approximation is an attractive method for modeling non-Gaussian distributions. Moreover, since GMs can be viewed as conditionally Gaussian distributions, they enable the analytic evaluation of the posterior. Consequently, they have been used for modeling the non-Gaussian noise distributions~\cite{bar2001estimation,GPFkotecha_gaussian_2003,chen_mixture_2000,alspach_nonlinear_1972,sorenson_recursive_1971,schoenberg_posterior_2012,kemouche_gmm_2010,bilik_mmse-based_2010,sun_mixture_2004,ali-loytty_box_2010,faubel_split_2009,wang_novel_2012,bilik_optimal_2005,daeipour_interacting_1995,pishdad_indoor_2012,pishdad_optimal_2013}, prior~\cite{anderson1979optimal,alspach_nonlinear_1972,sorenson_recursive_1971,flament_particle_2004,maybeck_multiple_2005,morelande_manoeuvring_2005,lehn-schioler_parzen_2004,bilik_optimal_2005}, likelihood~\cite{huber_efficient_2007}, and also the posterior~\cite{ito_gaussian_2000,GPFkotecha_gaussian_2003,schoenberg_posterior_2012,terejanu_adaptive_2011,bilik_maneuvering_2010,faubel_split_2009,wang_novel_2012,djuric_density_2004,schrempf_efficient_2007,bethel_pdf_2010}.

For linear dynamic systems with GM noise statistics, a bank of Kalman filters, or Gaussian Sum Filter (GSF), can be used to track the posterior and provide the MMSE state estimate~\cite{ackerson_state_1970,bar2001estimation}. However, the MMSE itself is not analytically tractable. Moreover, the general analytic bounds on MMSE including Posterior Cramer-Rao~\cite{tichavsky_posterior_1998}, Bobrovsky-Zakai~\cite{bobrovsky_lower_1976}, and Weiss-Weinstein~\cite{weiss_lower_1985,xaver_analytic_2013} become non-tractable for GM noise statistics. Thus, in this work we evaluate the MMSE and provide analytic lower and upper bounds that unlike~\cite{svensson_bayesian_2010,tulsyan_particle_2013}, do not require complex numerical approximations. This is particularly important, as the MMSE is the best achievable MSE of any estimator, and by having low complexity methods to evaluate its bounds, suitable filtering schemes for specific applications can be designed. This work can be considered an extension to~\cite{flam_mmse_2012}, in which the authors propose analytic bounds for static systems with GM noise distributions. Contrarily, in this work we consider dynamic systems which are more appropriate for mobile object positioning and location tracking. Moreover, we propose two implementable filters which have MSEs acting as upper bounds for the MMSE. The choice between these two upper bounds depends on the shape of the GM noise parameters. Moreover, since they are the MSEs of implementable filters, they can be used instead of the MMSE filter for applications where the upper bound for the MMSE provides the desired precision. The system and noise models used in this work, are the same models that we used in~\cite{pishdad_approximate_2014,pishdad_new_2014}. However, the purpose of this work is to provide analytic bounds on the MSE of the MMSE state estimator, rather than proposing new filtering or reduction schemes.

The rest of this paper is organized as follows: In Section~\ref{sec:Notations} we provide the details of the system model and introduce the notation used in this paper, as well as the details of GSF, which is the MMSE filter. In Section~\ref{sec:ContributionMSE}, we evaluate the MMSE and provide its analytic lower bound in Section~\ref{sec:LB}. In Section~\ref{sec:UB} we propose two upper bounds corresponding to the MSEs of two implementable filters. Depending on the shape of the GM noise models in the system, the tighter upper bound can be chosen. This is shown through simulations in Section~\ref{sec:SimulationResults}. Finally, the paper is concluded in Section~\ref{sec:Conclusion}.

\section{System Model}
\label{sec:Notations}
Suppose a discrete-time linear dynamic system, described by the following dynamics and measurement equations:
\begin{align}
\label{eq:process}
\mathbf{x}_{k} & = F_{k}\mathbf{x}_{k-1}+\mathbf{v}_{k},\\
\label{eq:measurement}
\mathbf{z}_k & = H_k\mathbf{x}_{k} + \mathbf{w}_k,
\end{align}
where \(\{\mathbf{x}_{k},k \in  \mathbb{N}\}\), and \(\{\mathbf{z}_{k},k \in  \mathbb{N}\}\) are the state and measurement sequences, respectively. The process noise, \(\{\mathbf{v}_{k},k \in  \mathbb{N}\}\) is an i.i.d. random vector sequence with the known pdf \(p{ \left(\mathbf{v}_{k} \right)}\).   The measurement noise, \(\{\mathbf{w}_{k},k \in  \mathbb{N}\}\) is also an i.i.d. random vector with the known pdf \(p{\left( \mathbf{w}_{k} \right)}\), and it is independent from the process noise. The matrices \(F_k\) and \(H_k\) are known and they define the linear relationships between the current and previous state vectors, and the current state and measurement vectors, respectively. \corcol{If we denote the size of the state vector by \(n_x\) and the size of the measurement vector by \(n_z\), the process noise is of size \(n_x\) and the measurement noise is of size \(n_z\). The matrices \(F_k\) and \(H_k\) are of size \(n_x \times n_x\) and \(n_z \times n_x\), respectively.}

In many signal processing applications, it is required to estimate the current unobservable state of the system, \(\mathbf{x}_k\), from its available noisy measurements, \(\mathbf{z}_{1:k}\). In positioning problems, the unobservable state denotes the unknown location, and the measurements are the information gathered from the location sensors. Therefore,~\eqref{eq:process} represents the motion model and~\eqref{eq:measurement} denotes the sensor model.

Since the MMSE estimator of state is the expected value of the posterior distribution~\cite{bar2001estimation,poor_introduction_1994}, in Bayesian tracking techniques the state is estimated probabilistically, by approximating the posterior distribution, \(p {\left(\mathbf{x}_k|\mathbf{z}_{1:k}\right)} \). For the special case of Gaussian process and measurement noise distributions, \corcol{the} Kalman filter optimally tracks the mean and covariance matrix of the Gaussian posterior~\cite{ho_bayesian_1964}, hence providing the MMSE state estimate. However, for non-Gaussian noise distributions other approximations need to be used. 

Gaussian Mixtures (GM) have been an attractive method for approximating non-Gaussian noise distributions, as they provide asymptotically unbiased estimations~\cite{1962KernelDensityEstimation}, with the desired level of accuracy~\cite{anderson1979optimal}.\footnote{The parameters can be chosen such that the integral of the approximation error over the sample space is as small as desired.} Moreover, by using the conditionally Gaussian representation for Gaussian Mixtures (GM), the MMSE state estimator can be evaluated by a bank of Kalman filters, or Gaussian Sum Filter (GSF)~\cite{bar2001estimation,ackerson_state_1970} (see Section~\ref{sec:GSF}). Therefore, in this work we assume that the process noise is estimated by a GM distribution\footnote{\corcol{Expectation maximization (EM) algorithm can be used to approximate a distribution by GMs, for instance see~\cite{mclachlan2004finite}.}} with \(C_{v_k}\) clusters, cluster means \(\left\lbrace \mathbf{u}^i_k,1 \leq i \leq C_{v_k}\right\rbrace  \), cluster covariance matrices \(\left\lbrace Q^i_k,1 \leq i \leq C_{v_k}\right\rbrace\), and mixing coefficients \(\left\lbrace \corcol{\mathcal{W}^i_k},1 \leq i \leq C_{v_k}\right\rbrace\), i.e. 
\begin{align}
\label{eq:priorDist}
p {\left(\mathbf{v}_k\right)} \approx \sum \limits _{i=1}^{C_{v_k}} \corcol{\mathcal{W}^i_k} \mathcal{N}{\left( \mathbf{v}_k; \mathbf{u}^i_k, Q^i_k \right)},
\end{align}
where \(\sum _{i}\corcol{\mathcal{W}^i_k}=1\), and \(\mathcal{N}{\left( \mathbf{x};\boldsymbol{\mu},\Sigma \right)} \) represents a Gaussian distribution with argument \(\mathbf{x}\), mean \(\boldsymbol{\mu}\), and covariance matrix \(\Sigma\). Similarly, the measurement noise distribution can also be approximated as:
\begin{align}
\label{eq:likelihoodDist}
p {\left(\mathbf{w}_k\right)} \approx \sum \limits _{j=1}^{C_{w_k}} \corcol{\mathcal{P}^j_k} \mathcal{N}{\left( \mathbf{w}_k; {\mathbf{b}}^j_k, R^j_k \right)},
\end{align}
where, \(C_{w_k}\) is the number of clusters of the GM distribution with coefficients \(\left\lbrace \corcol{\mathcal{P}^j_k},1 \leq j \leq C_{w_k}\right\rbrace  \) and \(\sum _{j} \corcol{\mathcal{P}^j_k}=1\). The mean and covariance matrix of cluster \( j,1 \leq j \leq C_{w_k}\) are \({\mathbf{b}}^j_k\) and \(R^j_k\), respectively. 

\subsection{MMSE Filter}
\label{sec:GSF}
Using the pdf approximations in~\eqref{eq:priorDist}--\eqref{eq:likelihoodDist} for the system defined in~\eqref{eq:process}--\eqref{eq:measurement}, the posterior can be viewed as having multiple \textit{models}, \(\left\lbrace M_{k}^{ij}; 1 \leq i \leq {C_{v_k}}, 1\leq j \leq {C_{w_k}} \right\rbrace\), each corresponding to a \textit{mode} in the posterior, which is a unique combination of the clusters in~\eqref{eq:priorDist} and \eqref{eq:likelihoodDist}. Hence, we can write:
\begin{align}
\label{eq:GSFmergedposterior}
p {\left(\mathbf{x}_k | \mathbf{z}_{1:k}\right)} & = \sum \limits _{i,j} \mu_k^{ij}{\left(\mathbf{z}_{k}\right)}p{\left(\mathbf{x}_k \big| \mathbf{z}_{1:k},M_{k}^{ij}\right)},
\end{align}
which denotes a GM distribution with \({C_{v_k}}\times{C_{w_k}}\) clusters, with coefficients
\begin{align}
\mu_k^{ij}{\left(\mathbf{z}_{k}\right)} & \triangleq  p {\left(M_{k}^{ij}\big|\mathbf{z}_{1:k}\right)}.
\end{align}
The \textit{mode-conditioned} posterior, \(p {\left(\mathbf{x}_k \big| \mathbf{z}_{1:k},M_{k}^{ij}\right)}\), is a Gaussian distribution with the pdf,
\begin{align}
\label{eq:individualKFposterior}
p {\left( \mathbf{x}_k \big|M_{k}^{ij},\mathbf{z}_{1:k} \right)} & = \mathcal{N} {\left(\mathbf{x}_k; \mathsf{\hat{\boldsymbol{x}}}_{k|k}^{ij}, \mathsf{P}_{k|k}^{ij}\right)},
\end{align}
and its parameters \(\mathsf{\hat{\boldsymbol{x}}}_{k|k}^{ij}\) and \(\mathsf{P}_{k|k}^{ij}\) can be evaluated using the \textit{mode-matched} Kalman filter~\cite{ho_bayesian_1964,bar2001estimation}, as follows:
\begin{align}
\label{eq:ijpredictedState}
\mathbf{\hat{x}}_{k|k-1}^{i} &= F_{k}\mathbf{\hat{x}}_{k-1|k-1} + \mathbf{u}_k^{i},
\\ \label{eq:ijpredictedCov} P_{k|k-1}^{i} & = Q_k^{i}+F_kP_{k-1|k-1}F_k\tran,
\\ \label{eq:ijpredictedMeans} \mathbf{\hat{z}}_{k}^{ij} &= H_k\mathbf{\hat{x}}_{k|k-1}^{i} + \mathbf{b}_k^{j},
\\ \label{eq:ijInnovationCov} S_k^{ij} & = H_k P_{k|k-1}^{i}H_k\tran + R_k^j,
\\ \label{eq:ijKalGain} \mathsf{W}_k^{ij} &= P_{k|k-1}^{i}H_k\tran{S_k^{ij}}^{-1},
\\ \label{eq:ijestimatedState} \mathsf{\hat{\boldsymbol{x}}}_{k|k}^{ij} &=\mathbf{\hat{{x}}}_{k|k-1}^{i} +\mathsf{W}_k^{ij} \left(\mathbf{z}_{k} - \mathbf{\hat{z}}_{k}^{ij}\right) ,
\\ \label{eq:ijKalstateCov} \mathsf{P}_{k|k}^{ij} & = P_{k|k-1}^{ij} -\mathsf{W}_k^{ij}S_k^{ij}\mathsf{W}{_k^{ij}}\tran,
\end{align}
where \(\mathbf{\hat{x}}_{k-1|k-1}\) and \(P_{k-1|k-1}\) are the  estimated state and the estimation error covariance matrix at \(k-1\), respectively, and \(\left(.\right)\tran\) indicates the transpose of its argument. \corcol{The matrix \(S_k^{ij}\) and the vector \(\mathsf{W}_k^{ij}\) represent the measurement prediction covariance matrix and the filter gain, respectively.}
Consequently, the posterior in~\eqref{eq:GSFmergedposterior} can be tracked by a bank of \({C_{v_k}}\times{C_{w_k}}\) mode-matched Kalman filters~\cite{anderson1979optimal,bar2001estimation}. Using
\begin{align}
p{\left(\mathbf{z}_k \big|M_{k}^{ij},\mathbf{z}_{1:k-1}\right)} & = \mathcal{N}{\left(\mathbf{z}_k; \mathbf{\hat{z}}_{k}^{ij}, S_k^{ij}\right)}, \\
p{\left(M_{k}^{ij} \big|\mathbf{z}_{1:k-1}\right)} & = \corcol{\mathcal{W}^i_k\mathcal{P}^j_k},
\end{align}
the coefficients \(\mu_k^{ij}{\left(\mathbf{z}_{k}\right)}\) are evaluated as:
\begin{align}
\mu_k^{ij}{\left(\mathbf{z}_{k}\right)} & = p{\left(M_{k}^{ij} \big|\mathbf{z}_k,\mathbf{z}_{1:k-1}\right)}
\\ & = \frac{p{\left(\mathbf{z}_k \big|M_{k}^{ij},\mathbf{z}_{1:k-1}\right)} p{\left(M_{k}^{ij} \big|\mathbf{z}_{1:k-1}\right)}}{p{\left(\mathbf{z}_{k}|\mathbf{z}_{1:k-1}\right)}}
\\ \label{eq:posteriorCoeff} & = \frac{\corcol{\mathcal{W}^i_k\mathcal{P}^j_k}\mathcal{N}{\left(\mathbf{z}_k ; \mathbf{\hat{z}}_{k}^{ij}, S_k^{ij}\right)}}{\sum \limits _{l,m} \corcol{\mathcal{W}^l_k\mathcal{P}^m_k}\mathcal{N}{\left(\mathbf{z}_k; \mathbf{\hat{z}}_{k}^{lm}, S_k^{lm}\right)}}.
\end{align}

Since the posterior density in~\eqref{eq:GSFmergedposterior} is a Gaussian mixture, having the mode-conditioned state estimations and covariance matrices, we can find the MMSE state estimate, 
\begin{eqnarray}
\label{eq:MMSEStateestimate}
\xmmse \triangleq \expectx {\left\lbrace \mathbf{{x}}_{k} |\mathbf{{z}}_{1:k} \right\rbrace }
= \sum \limits _{i,j} \mu_{k}^{ij}{\left(\mathbf{z}_{k}\right)}\mathsf{\hat{\boldsymbol{x}}}_{k|k}^{ij}, 
\end{eqnarray}
and the covariance matrix of the combined estimate,
\begin{eqnarray}
\label{eq:MMSECov} \Pmmse \triangleq  \expectx{\left\lbrace \left(\mathbf{x}_{k}-\hat{\mathbf{{x}}}_{k}^{\mathrm{MMSE}}\right)\left(\mathbf{x}_{k}-\hat{\mathbf{{x}}}_{k}^{\mathrm{MMSE}}\right)\tran \big| \mathbf{{z}}_{1:k} \right\rbrace }\\ \label{eq:MMSECov2}
=  \sum \limits_{i,j}\mu_{k}^{ij}{\left(\mathbf{z}_{k}\right)} \left( \mathsf{P}_{k|k}^{ij} + \mathsf{\hat{\boldsymbol{x}}}_{k|k}^{ij} \mathsf{\hat{\boldsymbol{x}}}{_{k|k}^{ij}}\tran \right)- \mathsf{\hat{\boldsymbol{x}}}_{k|k}\mathsf{\hat{\boldsymbol{x}}}_{k|k}\tran,
\end{eqnarray}
where \(\mathds{E}_\mathbf{x}{\left\lbrace g{\left(\mathbf{{x}}\right)} \right\rbrace} \) is the expected value of the function \(g{\left(\mathbf{{x}}\right)}\) with respect to the random variable \(\mathbf{{x}}\) with pdf \(p{\left(\mathbf{{x}}\right)}\)~\cite{bar2001estimation}\corcol{, and \(\mathsf{\hat{\boldsymbol{x}}}_{k|k} \triangleq \expectx {\left\lbrace \mathbf{{x}}_{k} |\mathbf{{z}}_{1:k} \right\rbrace }\)}.

\section{Unconditional MSE Evaluation and Bounds}
\label{sec:ContributionMSE}
Using an approach similar to~\cite{flam_mmse_2012}, in this section we evaluate the unconditional MSE of the MMSE filter, denoted by \(\epsilon_k ^2\).

\begin{lem}
\label{lemma:uncondMSE}
The unconditional MSE of a sequential state estimator, with the state estimation error covariance matrix \(P_{k|k}\), is equal to
\begin{align}
\label{eq:simplifiedUnconditionalMSE}
\int \mathrm{tr}{\left(P_{k|k}\right)}p{\left(\mathbf{{z}}_{k}|\mathbf{{z}}_{1:k-1}\right)} \mathrm{d}\mathbf{{z}}_{k},
\end{align}
where \(\mathrm{tr} {\left(.\right)}\) denotes the trace of its argument.
\end{lem}
\begin{IEEEproof}
See Appendix~\ref{sec:uncondMSEProof}.
\end{IEEEproof}
Using Lemma~\ref{lemma:uncondMSE}, and~\eqref{eq:MMSECov}, we can write:
\begin{align}
\label{eq:defMMSE}
\epsilon_k ^2 = \mathrm{tr}\left({\mathcal{M}_k}\right),
\end{align}
where
\begin{align}
\label{eq:defM}
\mathcal{M}_k \triangleq & \int \Pmmse p{\left(\mathbf{{z}}_{k}|\mathbf{{z}}_{1:k-1}\right)} \mathrm{d}\mathbf{{z}}_{k}.
\end{align} 
To evaluate the MMSE, we find the matrix \(\mathcal{M}_k\). Using~\eqref{eq:MMSECov2} we can write
\begin{align}
\label{eq:decomposeM}
\mathcal{M}_k =& \mathcal{M}_k^1+\mathcal{M}_k^2-\mathcal{M}_k^3,
\end{align}
where,
\begin{align}
\label{eq:defM1}
\mathcal{M}_k^1 \triangleq & \int \left(\sum \limits _{i,j}\mu_{k}^{ij}{\left(\mathbf{z}_{k}\right)}\mathsf{P}_{k|k}^{ij}p{\left(\mathbf{{z}}_{k}|\mathbf{{z}}_{1:k-1}\right)}\right) \mathrm{d}\mathbf{{z}}_{k} ,
\\ \label{eq:defM2}
\mathcal{M}_k^2 \triangleq & \int \left( \sum \limits _{i,j}\mu_{k}^{ij}{\left(\mathbf{z}_{k}\right)} \mathsf{\hat{\boldsymbol{x}}}_{k|k}^{ij} \mathsf{\hat{\boldsymbol{x}}}{_{k|k}^{ij}}\tran \right) p{\left(\mathbf{{z}}_{k}|\mathbf{{z}}_{1:k-1}\right)} \mathrm{d}\mathbf{{z}}_{k},\\ \label{eq:defM3}
\mathcal{M}_k^3 \triangleq & \int  \mathsf{\hat{\boldsymbol{x}}}_{k|k}\mathsf{\hat{\boldsymbol{x}}}_{k|k}\tran p{\left(\mathbf{{z}}_{k}|\mathbf{{z}}_{1:k-1}\right)} \mathrm{d}\mathbf{{z}}_{k}.
\end{align}
Additionally, from~\eqref{eq:posteriorCoeff}, we have
\begin{align}
\label{eq:muijOtherForm}
\mu_{k}^{ij}{\left(\mathbf{z}_{k}\right)} = & \frac{\corcol{\mathcal{W}^i_k\mathcal{P}^j_k}\mathcal{N}{\left(\mathbf{z}_k; \mathbf{\hat{z}}_{k}^{ij}, S_k^{ij}\right)}}{p{\left(\mathbf{{z}}_{k}|\mathbf{{z}}_{1:k-1}\right)}}.
\end{align}
Hence, \(\mathcal{M}_k^1\) is evaluated as
\begin{align}
\nonumber
\mathcal{M}_k^1 = & \sum \limits _{i,j} \corcol{\mathcal{W}^i_k\mathcal{P}^j_k} \int \frac{\mathcal{N}\left(\mathbf{z}_k; \mathbf{\hat{z}}_{k}^{ij}, S_k^{ij}\right)}{p{\left(\mathbf{{z}}_{k}|\mathbf{{z}}_{1:k-1}\right)}}\\
& \times \mathsf{P}_{k|k}^{ij}p{\left(\mathbf{{z}}_{k}|\mathbf{{z}}_{1:k-1}\right)}\mathrm{d}\mathbf{{z}}_{k} 
\\ \label{eq:M1} = &
 \sum \limits _{i,j} \corcol{\mathcal{W}^i_k\mathcal{P}^j_k} \mathsf{P}_{k|k}^{ij},
\end{align}
since \(\mathsf{P}_{k|k}^{ij}\) is not a function of the measurements (see~\eqref{eq:ijKalstateCov}). 

Now, using~\eqref{eq:muijOtherForm},~\eqref{eq:ijestimatedState}, and~\eqref{eq:ijKalstateCov} we can write:
\begin{align}
\nonumber
\mathcal{M}_k^2 = & \sum \limits _{i,j} \corcol{\mathcal{W}^i_k\mathcal{P}^j_k} \int \left(\mathbf{\hat{{x}}}_{k|k-1}^{i} +\mathsf{W}_k^{ij} \left( \mathbf{z}_{k} - \mathbf{\hat{z}}_{k}^{ij} \right)\right)\\
\nonumber
& \times \left(\mathbf{\hat{{x}}}_{k|k-1}^{i} 
 +\mathsf{W}_k^{ij} \left( \mathbf{z}_{k} - \mathbf{\hat{z}}_{k}^{ij} \right)\right)\tran \\ 
& \times {\mathcal{N}{\left(\mathbf{z}_k; \mathbf{\hat{z}}_{k}^{ij}, S_k^{ij}\right)}}\mathrm{d}\mathbf{{z}}_{k}
\\ \label{eq:M2} = & \sum \limits _{i,j} \corcol{\mathcal{W}^i_k\mathcal{P}^j_k} \left(\mathbf{\hat{{x}}}_{k|k-1}^{i}\mathbf{\hat{{x}}}{_{k|k-1}^{i}}\tran + \mathsf{W}_k^{ij}S_k^{ij} \mathsf{W}{_k^{ij}}\tran \right).
\end{align}

To evaluate \(\mathcal{M}_k^3\), we use~\eqref{eq:MMSEStateestimate} and write:
\begin{align}
\nonumber
\mathcal{M}_k^3 = & \int \sum \limits _{i,j} \sum \limits _{l,m} \mu_{k}^{ij}{\left(\mathbf{z}_{k}\right)} \mu_{k}^{lm}{\left(\mathbf{z}_{k}\right)}\mathsf{\hat{\boldsymbol{x}}}_{k|k}^{ij} \mathsf{\hat{\boldsymbol{x}}}{_{k|k}^{lm}}\tran p{\left(\mathbf{{z}}_{k}|\mathbf{{z}}_{1:k-1}\right)} \mathrm{d}\mathbf{{z}}_{k}
\\ 
\nonumber
= & \sum \limits _{i,j} \sum \limits _{l,m}\corcol{\mathcal{W}^i_k\mathcal{P}^j_k}\corcol{\mathcal{W}^l_k\mathcal{P}^m_k} \int \mathcal{N}{\left(\mathbf{z}_k; \mathbf{\hat{z}}_{k}^{ij}, S_k^{ij}\right)}\\
& \times \frac{ \mathcal{N}{\left(\mathbf{z}_k; \mathbf{\hat{z}}_{k}^{lm}, S_k^{lm}\right)}}{\sum \limits _{r,s} \corcol{\mathcal{W}^r_k\mathcal{P}^s_k}\mathcal{N}{\left(\mathbf{z}_k; \mathbf{\hat{z}}_{k}^{rs}, S_k^{rs}\right)}}  \mathsf{\hat{\boldsymbol{x}}}_{k|k}^{ij} \mathsf{\hat{\boldsymbol{x}}}{_{k|k}^{lm}}\tran \mathrm{d}\mathbf{{z}}_{k},
\end{align}
which cannot be evaluated in closed-form due to the GM pdf in the denominator. Additionally, since it cannot be written as an expected value, numerical methods used for evaluating expectations cannot be applied to this problem. Hence, in this work we propose bounds for the MMSE. 
%

\subsection{Lower Bound for the MMSE}
\label{sec:LB}
The MMSE filter evaluates and tracks the mode-conditioned posteriors and combines them using~\eqref{eq:MMSEStateestimate} to provide the state estimation. However, at each iteration only one of the models is \textit{active}, corresponding to the process and measurement noise clusters in effect at that iteration. Consequently, if prior information about the active model is available, by only including the posterior cluster corresponding to this model in the estimation of the current state, the total MSE is minimized. This is due to the fact that Kalman filter is the optimal state estimator if the model parameters are selected correctly~\cite{arulampalam_tutorial_2002,ho_bayesian_1964}. Assuming that cluster \(i\) and \(j\) are the active clusters of the process and measurement noise, respectively, we denote the current active model by \(M^{ij*}_k\). The \textit{Matched filter} state estimation and covariance matrix can then be defined as:
\begin{align}
\hat{\mathbf{{x}}}_{k}^{*} \triangleq \mathsf{\hat{\boldsymbol{x}}}_{k|k}^{ij}, \;
P_{k|k}^* \triangleq  \mathsf{P}_{k|k}^{ij}.
\end{align}
Hence, we have\footnote{It is worth noting that there could exist another model, \(M^{lm}_k\neq M^{ij*}_k\), such that
\begin{align}
\left\Vert\mathbf{x}_{k}-\mathsf{\hat{\boldsymbol{x}}}_{k|k}^{lm}\right\Vert_{2} \leq \left\Vert\mathbf{x}_{k}-\hat{\mathbf{{x}}}_{k}^{*}\right\Vert_2,
\end{align}
where \(\left\Vert.\right\Vert_{p}\) indicates the \(p\)-norm of its argument.
Hence, the MSE of the estimator \(\mathsf{\hat{\boldsymbol{x}}}_{k|k}^{lm}\) will be lower than \(P_{k|k}^*\). However, since the active model is \(M^{ij*}_k\), this estimator can be biased. }
\begin{align}
\label{eq:LBuneq}
\Pmmse \geq P_{k|k}^*.
\end{align}
Now if \(\epsilon_k ^{*2}\) denotes the MSE of Matched filter, using Lemma~\ref{lemma:uncondMSE}, we can write
\begin{align}
\label{eq:defMatchedMSE}
\epsilon_k ^{*2} = & \int \mathrm{tr}{\left(P_{k|k}^*\right)}p{\left(\mathbf{{z}}_{k}|\mathbf{{z}}_{1:k-1}\right)} \mathrm{d}\mathbf{{z}}_{k}.
\end{align}
Using~\eqref{eq:defM},~\eqref{eq:LBuneq}, and~\eqref{eq:defMatchedMSE}, we have
\begin{align}
\epsilon_k ^{2} \geq \mathrm{tr}{\left(\mathcal{M}^*_k\right)}.
\end{align}
To evaluate \(\epsilon_k ^{*2}\), we define
\begin{align}
\mathcal{M}^*_k \triangleq & \int P_{k|k}^*p{\left(\mathbf{{z}}_{k}|\mathbf{{z}}_{1:k-1}\right)} \mathrm{d}\mathbf{{z}}_{k} \\
= & \int \sum \limits _{i,j} p{\left(M_{k}^{ij}|\mathbf{z}_{1:k}\right)} \mathsf{P}_{k|k}^{ij}p{\left(\mathbf{{z}}_{k}|\mathbf{{z}}_{1:k-1}\right)} \mathrm{d}\mathbf{{z}}_{k} \\
= & \int \sum \limits _{i,j} \mu_{k}^{ij}{\left(\mathbf{z}_{k}\right)}\mathsf{P}_{k|k}^{ij}p{\left(\mathbf{{z}}_{k}|\mathbf{{z}}_{1:k-1}\right)} \mathrm{d}\mathbf{{z}}_{k}\\
\label{eq:LBMat} = &
 \sum \limits _{i,j} \corcol{\mathcal{W}^i_k\mathcal{P}^j_k} \mathsf{P}_{k|k}^{ij}.
\end{align}
\begin{corol}
\label{corolLB}
The lower bound for the MMSE, is the unconditional MSE of Matched filter, i.e.
\begin{align}
\label{eq:LB}
\epsilon_k ^{2} \geq \sum \limits _{i,j} \corcol{\mathcal{W}^i_k\mathcal{P}^j_k} \mathrm{tr}{\left(\mathsf{P}_{k|k}^{ij}\right)}.
\end{align}
\end{corol}
Alternatively, Corollary~\ref{corolLB} can be proved using an approach similar to~\cite{flam_mmse_2012}. Specifically, using the definitions in~\eqref{eq:defM1}--\eqref{eq:defM3}, we have
\begin{align}
\mathrm{tr}{\left(\mathcal{M}_k^1\right)}, \mathrm{tr}{\left(\mathcal{M}_k^2\right)}, \mathrm{tr}{\left(\mathcal{M}_k^3\right)} \geq 0.
\end{align}
Additionally, 
\begin{align}
\mathrm{tr}{\left(\mathcal{M}_k^2\right)}-\mathrm{tr}{\left(\mathcal{M}_k^3\right)} \geq 0,
\end{align}
since it is the expected value of the spread of means,
\begin{align} 
\sum \limits _{i,j} \mu_{k}^{ij}{\left(\mathbf{z}_{k}\right)}\left( \mathsf{\hat{\boldsymbol{x}}}_{k|k}^{ij}- \mathsf{\hat{\boldsymbol{x}}}_{k|k}\right)\left( \mathsf{\hat{\boldsymbol{x}}}_{k|k}^{ij}- \mathsf{\hat{\boldsymbol{x}}}_{k|k}\right)^T.
\end{align}
Hence, we have
\begin{align}
\label{eq:alternateLBproof}
\epsilon_k ^{2} \geq \mathrm{tr}{\left(\mathcal{M}^1_k\right)}.
\end{align}
By applying~\eqref{eq:M1} to~\eqref{eq:alternateLBproof},~\eqref{eq:LB} is proved. \ \ \ \ \ \ \ \ \ \ \ \ \ \ \ \ \ \ \ \ \ \ \ \
\IEEEQED

\subsection{Upper Bound for the MMSE}
\label{sec:UB}
Since the Matched filter is not practical,\footnote{Due to the unavailability of prior information about the active mode.} the model corresponding to the posterior cluster with the maximum weight, \(\mu_k^{ij}{\left(\mathbf{z}_{k}\right)}\) can be selected as the active model. In this section, we show that this filter provides an upper bound for the unconditional MSE of the MMSE filter.

\begin{Def}
\label{def:UBFilter}
Assuming a GSF where cluster \(ij\) in the posterior has the maximum weight, i.e.
\begin{align}
\label{eq:determineActive}
\mu_k^{ij}{\left(\mathbf{z}_{k}\right)} = \max_{l,m}\mu_k^{lm}{\left(\mathbf{z}_{k}\right)},
\end{align}
GSF-R is defined as the Kalman filter \(ij\) in GSF. In other words, GSF-R selects and uses the Kalman filter with the maximum weight in GSF. We also define
\begin{align}
M_k^{\mathrm{R}} \triangleq  M_k^{ij}, \;
\hat{\mathbf{x}}_{k|k}^{\mathrm{R}} \triangleq  \mathsf{\hat{\boldsymbol{x}}}_{k|k}^{ij}, \;
 P_{k|k}^{\mathrm{R}}  \triangleq  \mathsf{P}_{k|k}^{ij},
\end{align}
to address the parameters of GSF-R, when~\eqref{eq:determineActive} is true. \corcol{It is worth noting that only if \(M^{ij*}_k=M_k^{\mathrm{R}}\), \(P_{k|k}^{\mathrm{R}}\) represents the true covariance matrix of this filter. Therefore, we define}
\corcol{\begin{align}
C_{k|k}^{\mathrm{R}} \triangleq \expectx{\left\lbrace \left(\mathbf{x}_{k}-\hat{\mathbf{{x}}}_{k}^{\mathrm{R}}\right)\left(\mathbf{x}_{k}-\hat{\mathbf{{x}}}_{k}^{\mathrm{R}}\right)\tran \big| \mathbf{{z}}_{1:k} \right\rbrace }.
\end{align}}
\end{Def}

\begin{Prop}
\label{prop:UB}
The unconditional MSE of GSF-R, acts as an upper bound for the MMSE.
\end{Prop}
\begin{IEEEproof}
Using~\eqref{eq:determineActive}, \(M_k^{ij}\) will be selected as \(M_k^{R}\) if \(\mathbf{z}_k\) is in \(\mathcal{R}^{ij}_k \), where
\begin{eqnarray}
\nonumber
\mathcal{R}^{ij}_k \triangleq \Big \lbrace \mathbf{z}_k: \forall \; lm\neq ij; 1 \leq i,l \leq {C_{v_k}}, 1\leq j,m \leq C_{w_k}; \Big. \\ \label{eq:defRij}
\Big. \corcol{\mathcal{W}^i_k\mathcal{P}^j_k}\mathcal{N}{\left(\mathbf{z}_k ; \mathbf{\hat{z}}_{k}^{ij}, S_k^{ij}\right)} > \corcol{\mathcal{W}^l_k\mathcal{P}^m_k}\mathcal{N}{\left(\mathbf{z}_k ; \mathbf{\hat{z}}_{k}^{lm}, S_k^{lm}\right)}\Big\rbrace.
\end{eqnarray}
Hence, we can write
\begin{align}
\label{eq:probijMax}
p{\left(M^{\mathrm{R}}_k = M_k^{ij}|\mathbf{z}_{1:k}\right)} = \mathds{1}_{\mathcal{R}^{ij}_k}{\left(\mathbf{z}_k\right)},
\end{align}
where \(\mathds{1}_\mathcal{A}{\left(.\right)}\) is the indicator function of set \(\mathcal{A}\). Thus, given the available measurements, the probability of selecting \(M^{ij}_k\) when the true active model is \(M^{lm*}_k\) is
\corcol{\begin{align}
\label{eq:probmaxijbutlm}
\gamma_k^{ij,lm*}{\left(\mathbf{z}_k\right)}  \triangleq & p{\left(M^{R}_k = M_k^{ij}, M_k^{lm*}\big|\mathbf{z}_{1:k}\right)}
\\
\label{eq:gammaijlm}
 = & \mathds{1}_{\mathcal{R}^{ij}_k}{\left(\mathbf{z}_k\right)}\mu_k^{lm}{\left(\mathbf{z}_{k}\right)} 
\end{align}
which is a function of the current observation and the parameters of the GM posterior, i.e. the coefficients, cluster means, and covariance matrices.}
Now, if \(\epsilon_k^{\mathrm{R}^2}\) denotes the MSE of GSF-R, using Lemma~\ref{lemma:uncondMSE} we have
\begin{align}
\label{eq:defUBMSE}
\epsilon_k ^{\mathrm{R}^2} = & \int \mathrm{tr}{\left(\corcol{C}_{k|k}^{\mathrm{R}}\right)}p{\left(\mathbf{{z}}_{k}|\mathbf{{z}}_{1:k-1}\right)} \mathrm{d}\mathbf{{z}}_{k},
\end{align}
which is equal to the trace of \(\mathcal{M}^{\mathrm{R}}\), where
\begin{align}
\label{eq:MUB}
\mathcal{M}^{\mathrm{R}}_k \triangleq &  \int \left(\corcol{C}_{k|k}^{\mathrm{R}}\right)p{\left(\mathbf{{z}}_{k}|\mathbf{{z}}_{1:k-1}\right)} \mathrm{d}\mathbf{{z}}_{k}.
\end{align}
The covariance matrix of GSF-R can be written as
\corcol{\begin{align}
\label{eq:CRemoveTermByTerm}
C_{k|k}^{\mathrm{R}} = & \expectx{\left\lbrace \mathbf{x}_{k}\mathbf{x}_{k}\tran - \mathbf{x}_{k}\hat{\mathbf{{x}}}{_{k}^{\mathrm{R}}}\tran  -\hat{\mathbf{{x}}}{_{k}^{\mathrm{R}}}\mathbf{x}_{k}\tran + \hat{\mathbf{{x}}}{_{k}^{\mathrm{R}}}\hat{\mathbf{{x}}}{_{k}^{\mathrm{R}}}\tran \big| \mathbf{{z}}_{1:k} \right\rbrace}.
\end{align}
The first term in~\eqref{eq:CRemoveTermByTerm}, can be evaluated using~\eqref{eq:MMSECov} as
\begin{align}
\expectx{\left\lbrace \mathbf{x}_{k}\mathbf{x}_{k}\tran | \mathbf{{z}}_{1:k}\right\rbrace} = \Pmmse + \mathsf{\hat{\boldsymbol{x}}}_{k|k}\mathsf{\hat{\boldsymbol{x}}}_{k|k}\tran.
\end{align}
Now, by conditioning on the active model and the selected model by GSF-R and using~\eqref{eq:gammaijlm}, we have
\begin{align}
\nonumber
\expectx{\left\lbrace \mathbf{x}_{k}\hat{\mathbf{{x}}}{_{k}^{\mathrm{R}}}\tran \big| \mathbf{{z}}_{1:k} \right\rbrace} = & \sum\limits_{i,j} \sum\limits_{l,m} \gamma_k^{ij,lm*}{\left(\mathbf{z}_k\right)}\\
& \expectx{\left\lbrace \mathbf{x}_{k}\hat{\mathbf{{x}}}{_{k}^{\mathrm{R}}}\tran \big|M^{R}_k = M_k^{ij}, M_k^{lm*}, \mathbf{{z}}_{1:k} \right\rbrace} \\
=& \sum\limits_{i,j} \sum\limits_{l,m} \mathds{1}_{\mathcal{R}^{ij}_k}{\left(\mathbf{z}_k\right)}\mu_k^{lm}{\left(\mathbf{z}_{k}\right)} \mathsf{\hat{\boldsymbol{x}}}_{k|k}^{lm}\mathsf{\hat{\boldsymbol{x}}}{_{k|k}^{ij}}\tran \\
=& \mathsf{\hat{\boldsymbol{x}}}_{k|k} \sum\limits_{i,j} \mathds{1}_{\mathcal{R}^{ij}_k}{\left(\mathbf{z}_k\right)} \mathsf{\hat{\boldsymbol{x}}}{_{k|k}^{ij}}\tran.
\end{align}
Similarly, we can find 
\begin{align}
\expectx{\left\lbrace \mathbf{x}_{k}\hat{\mathbf{{x}}}{_{k}^{\mathrm{R}}}\tran \big| \mathbf{{z}}_{1:k} \right\rbrace} = & \sum\limits_{i,j} \mathds{1}_{\mathcal{R}^{ij}_k}{\left(\mathbf{z}_k\right)} \mathsf{\hat{\boldsymbol{x}}}{_{k|k}^{ij}} \mathsf{\hat{\boldsymbol{x}}}_{k|k} \tran.
\end{align}
Finally, by conditioning on the selected model by GSF-R and using~\eqref{eq:probijMax}, we can find
\begin{align}
\nonumber
\expectx{\left\lbrace \hat{\mathbf{{x}}}{_{k}^{\mathrm{R}}}\hat{\mathbf{{x}}}{_{k}^{\mathrm{R}}}\tran \big| \mathbf{{z}}_{1:k} \right\rbrace} = & \sum\limits_{i,j}   \mathds{1}_{\mathcal{R}^{ij}_k}{\left(\mathbf{z}_k\right)}\\
& \expectx{\left\lbrace \hat{\mathbf{{x}}}{_{k}^{\mathrm{R}}}\hat{\mathbf{{x}}}{_{k}^{\mathrm{R}}}\tran \big|M^{R}_k = M_k^{ij}, \mathbf{{z}}_{1:k} \right\rbrace} \\
=& \sum\limits_{i,j}  \mathds{1}_{\mathcal{R}^{ij}_k} \mathsf{\hat{\boldsymbol{x}}}{_{k|k}^{ij}}\mathsf{\hat{\boldsymbol{x}}}{_{k|k}^{ij}}\tran.
\end{align}
Hence, \(C_{k|k}^{\mathrm{R}}\) can be written as
\begin{align}
\nonumber
C_{k|k}^{\mathrm{R}} =& \Pmmse + \mathsf{\hat{\boldsymbol{x}}}_{k|k}\mathsf{\hat{\boldsymbol{x}}}_{k|k}\tran - \mathsf{\hat{\boldsymbol{x}}}_{k|k} \sum\limits_{i,j} \mathds{1}_{\mathcal{R}^{ij}_k}{\left(\mathbf{z}_k\right)} \mathsf{\hat{\boldsymbol{x}}}{_{k|k}^{ij}}\tran \\ \label{eq:defCGSFR}
&- \sum\limits_{i,j} \mathds{1}_{\mathcal{R}^{ij}_k}{\left(\mathbf{z}_k\right)} \mathsf{\hat{\boldsymbol{x}}}{_{k|k}^{ij}} \mathsf{\hat{\boldsymbol{x}}}_{k|k} \tran +\sum\limits_{i,j}  \mathds{1}_{\mathcal{R}^{ij}_k}{\left(\mathbf{z}_k\right)} \mathsf{\hat{\boldsymbol{x}}}{_{k|k}^{ij}}\mathsf{\hat{\boldsymbol{x}}}{_{k|k}^{ij}}\tran\\
\nonumber 
=& \Pmmse 
+ \left(\mathsf{\hat{\boldsymbol{x}}}_{k|k} - \sum\limits_{i,j} \mathds{1}_{\mathcal{R}^{ij}_k}{\left(\mathbf{z}_k\right)} \mathsf{\hat{\boldsymbol{x}}}{_{k|k}^{ij}}\right)\\
\label{eq:defCGSFR2}
& \times \left(\mathsf{\hat{\boldsymbol{x}}}_{k|k} - \sum\limits_{i,j} \mathds{1}_{\mathcal{R}^{ij}_k}{\left(\mathbf{z}_k\right)} \mathsf{\hat{\boldsymbol{x}}}{_{k|k}^{ij}}\right)\tran,
\end{align}
as we have
\begin{align}
\nonumber
\sum\limits_{i,j}  \mathds{1}_{\mathcal{R}^{ij}_k}{\left(\mathbf{z}_k\right)} \mathsf{\hat{\boldsymbol{x}}}{_{k|k}^{ij}}\mathsf{\hat{\boldsymbol{x}}}{_{k|k}^{ij}}\tran = &
\left(\sum\limits_{i,j} \mathds{1}_{\mathcal{R}^{ij}_k}{\left(\mathbf{z}_k\right)}\mathsf{\hat{\boldsymbol{x}}}{_{k|k}^{ij}}\right) \\ 
& \times \left(\sum\limits_{i,j} \mathds{1}_{\mathcal{R}^{ij}_k}{\left(\mathbf{z}_k\right)}\mathsf{\hat{\boldsymbol{x}}}{_{k|k}^{ij}}\right)\tran.
\end{align}
Thus, 
\begin{align}
\Pmmse \leq C_{k|k}^{\mathrm{R}},
\end{align}
and using~\eqref{eq:defMMSE},~\eqref{eq:defM},~\eqref{eq:defUBMSE} we have}
\begin{align}
\label{eq:UB}
\epsilon_k ^{2} \leq \mathrm{tr}{\left(\mathcal{M}^{\mathrm{R}}_k\right)}. \ \IEEEQEDhere
\end{align}
\end{IEEEproof}
The details of evaluating \(\mathcal{M}^{\mathrm{R}}_k\), are provided in Appendix~\ref{sec:MUB}.

\subsection{Alternative Upper Bound}
Kalman filter provides the MMSE state estimation for Gaussian noise distributions by tracking the sufficient statistics of the Gaussian posterior. For non-Gaussian posteriors, however, Kalman filter can only provide the Linear MMSE (LMMSE) state estimation~\cite{poor_introduction_1994}. In this section, using an approach similar to~\cite{flam_mmse_2012}, we use the Linear MMSE (LMMSE), to find an upper bound for MMSE. 

To apply Kalman filter to the system defined in~\eqref{eq:process}--\eqref{eq:measurement}, we need to use the moment-matched Gaussian distributions of the GM noise pdfs in~\eqref{eq:priorDist}--\eqref{eq:likelihoodDist}. Hence, we define:
\begin{align}
\bar{\mathbf{u}}_k & \triangleq \sum \limits _{i} \corcol{\mathcal{W}^i_k}\mathbf{u}^i_k, \bar{\mathbf{b}}_k  \triangleq \sum \limits _{j} \corcol{\mathcal{P}^j_k}\mathbf{b}^j_k,
\\
 \bar{Q}_k & \triangleq \sum \limits _{i} \corcol{\mathcal{W}^i_k}  \left(Q^i_k + \left(\bar{\mathbf{u}}_k-\mathbf{u}^i_k\right)\left(\bar{\mathbf{u}}_k-\mathbf{u}^i_k\right)\tran\right),
\\
 \bar{R}_k & \triangleq \sum \limits _{j} \corcol{\mathcal{P}^j_k} \left(R^j_k + \left(\bar{\mathbf{b}}_k-\mathbf{b}^j_k\right)\left(\bar{\mathbf{b}}_k-\mathbf{b}^j_k\right)\tran\right).
\end{align}
Thus, the LMMSE can be evaluated by running a Kalman filter on a Gaussian process noise with mean \(\bar{\mathbf{u}}_k\), and covariance matrix \(\bar{Q}_k\), and a Gaussian measurement noise with mean \(\bar{\mathbf{b}}_k\) and covariance matrix \(\bar{R}_k\), as follows:
\begin{align}
P_{k|k-1}^{\mathrm{K}} = & F_kP_{k-1|k-1}F_k\tran + \bar{Q}_k ,\\
S_k^{\mathrm{K}} = & H_kP_{k|k-1}^{\mathrm{K}}H_k\tran + \bar{R}_k,\\
\label{eq:PKalman}
P_{k|k}^{\mathrm{K}} = & P_{k|k-1}^{\mathrm{K}} - P_{k|k-1}^{\mathrm{K}} H_k \tran {S_k^{\mathrm{K}}}^{-1}H_kP_{k|k-1}^{\mathrm{K}}.
\end{align}

\begin{Prop}
\label{prop:KalUL}
The MSE of LMMSE filter, acts as an upper bound for MMSE, i.e.
\begin{align}
\epsilon_k ^{2} \leq \mathrm{tr}{\left(P_{k|k}^{\mathrm{K}}\right)}.
\end{align}
\end{Prop}
\begin{IEEEproof}
Since the MMSE filter minimizes the MSE, any implementable filter will have an MSE greater or equal to the MSE of the MMSE filter.
\end{IEEEproof}
\begin{corol}
Upper bound of MMSE is the smaller of LMMSE and the MSE of GSF-R, i.e.
\begin{align}
\label{eq:CombinedUL}
\epsilon_k ^{2} \leq \min{\left\lbrace\mathrm{tr}{\left(\mathcal{M}^{\mathrm{R}}_k\right)}, \mathrm{tr}{\left(P_{k|k}^{\mathrm{K}}\right)}\right\rbrace}.
\end{align}
\end{corol}
The shape of the GM noise parameters can affect the choice of the upper bound. Specifically, increasing the distance between the means of clusters in the GM noise distributions, increases the distance between the clusters of the GM posterior distribution. Consequently, \(\mathrm{tr}{\left(P_{k|k}^{\mathrm{K}}\right)}\) will have a larger value. This is supported through simulations in Section~\ref{sec:SimulationResults}. However, this trend is not valid for \(\mathrm{tr}{\left(P_{k|k}^{\mathrm{R}}\right)}\), hence it is a better upper bound for highly multimodal GM posteriors (See Proposition~\ref{prop:ConvergenceofBounds}).
\begin{Prop}
\label{prop:ConvergenceofBounds}
Increasing the Mahalanobis distance between the clusters of the GM posterior decreases the difference between the bounds in~\eqref{eq:LB},~\eqref{eq:UB} and the MMSE. In the limit, the upper bound in~\eqref{eq:UB} and the MMSE, converge to the lower bound in~\eqref{eq:LB}.
\end{Prop}
\begin{IEEEproof}
Larger Mahalanobis distance among the clusters of the GM posterior, leads to increased difference between the likelihoods,
\(\left\lbrace\mathcal{N}\left(\mathbf{z}_k; \mathbf{\hat{z}}_{k}^{ij}, S_k^{ij}\right); 1 \leq i \leq {C_{v_k}}, 1\leq j \leq {C_{w_k}} \right\rbrace\).
In the limit, we have only one mixand corresponding to \(M^{ij*}\) for which \(\mathcal{N}\left(\mathbf{z}_k; \mathbf{\hat{z}}_{k}^{ij*}, S_k^{ij*}\right)=1\), thus \(\mu_{k}^{ij*}{\left(\mathbf{z}_{k}\right)} =  1\). Therefore, 
\begin{align}
\label{eq:muconverge2}
\forall lm \neq ij; \; \mu_{k}^{lm}{\left(\mathbf{z}_{k}\right)} =  0.
\end{align}
Hence, using~\eqref{eq:muconverge2} in~\eqref{eq:defM2},~\eqref{eq:defM3}, we have
\begin{align}
\mathrm{tr}{\left(\mathcal{M}_k^2\right)}-\mathrm{tr}{\left(\mathcal{M}_k^3\right)} =0,
\end{align}
which yields
\begin{align}
\label{eq:MSEConvergesLB}
\epsilon ^ 2 = 
 \sum \limits _{i,j} \corcol{\mathcal{W}^i_k\mathcal{P}^j_k} \mathsf{P}_{k|k}^{ij}.
\end{align}
Additionally, since in the limit there is no overlap between the mixands of the GM posterior, from~\eqref{eq:probmaxijbutlm}, we have
\begin{align}
\label{eq:gamaconverge}
\forall lm \neq ij; \; \gamma_k^{ij,lm*}{\corcol{\left({\mathbf{z}_k}\right)}} = 0.
\end{align}
Also, from\corcol{~\eqref{eq:defCGSFR2}}
\begin{align}
\label{eq:Dconverge}
\corcol{\corcol{C}_{k|k}^{\mathrm{R}} = \Pmmse}
\end{align}
\corcol{Using\eqref{eq:Dconverge} in~\eqref{eq:MSEConvergesLB}, we have}
\begin{align}
\corcol{\epsilon_k^{\mathrm{R}^2}} = & \sum\limits_{l,m} \corcol{\mathcal{W}^l_k\mathcal{P}^m_k} \mathsf{P}_{k|k}^{lm}. \ \IEEEQEDhere
\end{align}
\end{IEEEproof}

\section{Simulation Results}
\label{sec:SimulationResults}
In this section we use synthetically generated data to approximate the MMSE and compare it with the proposed bounds\corcol{, as well as the Posterior Cramer-Rao lower bound~\cite{tichavsky_posterior_1998}}. In our simulation scenario, we assume a positioning system, with the following parameters:
\begin{align}
F_k =  \begin{bmatrix}
1 & \Delta t_{k} \\
0 & 1
\end{bmatrix},
H_k = \begin{bmatrix}
1 & 0
\end{bmatrix},
\end{align}
where \(\Delta t_{k}\) is the time interval between the measurements \(z_{k-1}\) and \(z_k\).\footnote{We use scalar measurements.} In our simulations, we use \(\Delta t_k=0.1080\) s for all iterations, \(k\).

The process model used in our simulations, is a random walk velocity motion model. Therefore, in~\eqref{eq:process} we have:
\begin{align}
\mathbf{v}_{k} = v_k \times \begin{bmatrix}
\Delta t_{k} ,
1
\end{bmatrix}\tran,
\end{align}
where \(v_k\) is a univariate random variable. In our simulations, we use the same model\footnote{But with different separations among the clusters.} to generate process and measurement noise samples. To ensure that our simulation results are not model dependent, we used three different GM models to generate our data:
\begin{itemize}
{\setlength\itemindent{35pt}\item[\textbf{Model 1:}] Symmetric distribution with the mixands all having the same weights.}
{\setlength\itemindent{35pt}\item[\textbf{Model 2:}] Symmetric distribution with the mixands having not necessarily equal weights.}
{\setlength\itemindent{35pt}\item[\textbf{Model 3:}] Asymmetric distribution.}
\end{itemize}
The coefficients of the mixands are chosen such that the noise processes are zero mean. In our simulations, we assumed GM distributions with 5 clusters, each having a variance of \(1\). The coefficients and the means of the clusters are given in Table~\ref{tab:GMModelsparams}, and they are denoted by \(\mathbf{w}\) and \(\mathbf{m}\), respectively. The parameter \(c\) in this table is multiplied by the means to change the distance among the clusters of measurement noise distribution, i.e. the separation among the clusters. For the process noise, we use \(c=1\), to keep the state vector, hence the error, bounded.\footnote{As shown in~\cite{bar2001estimation}, if the filter is not completely consistent with the underlying system model, the estimation error is a function of the state vector. Therefore, if the state vector is unbounded, the filter will be unstable. To avoid this, we kept the variance of the state vector constant and only varied the measurement noise distribution.}
\begin{table}[!t]
\centering
\caption{The parameters of the GM models used for generating synthetic data}
\label{tab:GMModelsparams}
\begin{tabular}{c l}
\hline
\textbf{Model 1} & \(\mathbf{w} = \left[0.2,0.2,0.2,0.2,0.2\right]\) \\
& \(\mathbf{m}=c \left[-50,-30,0,30,50\right]\) \\
\hline
\textbf{Model 2} & \(\mathbf{w} = \left[0.1,0.1,0.6,0.1,0.1\right]\) \\
& \(\mathbf{m}=c\left[-50,-30,0,30,50\right]\) \\
\hline
\textbf{Model 3} & \(\mathbf{w} = \left[0.5,0.1,0.1,0.1,0.2\right]\) \\
& \(\mathbf{m}=c\left[-50,10,30,50,80\right]\) \\
\hline
\end{tabular}
\end{table}

\begin{figure}[!t]
\centering
\includegraphics[width=3.5in]{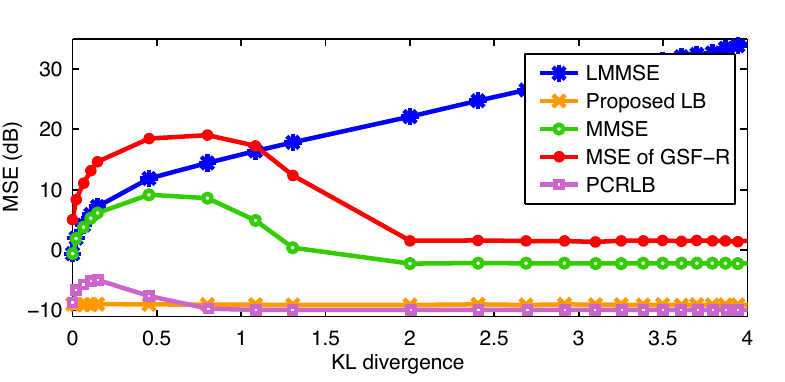}
\caption{MSE for the synthetic data generated using Model 1 vs. KL divergence between the measurement noise pdf and its moment-matched Gaussian.}
\label{fig:MSEM1}
\end{figure}

\begin{figure}[!t]
\centering
\includegraphics[width=3.5in]{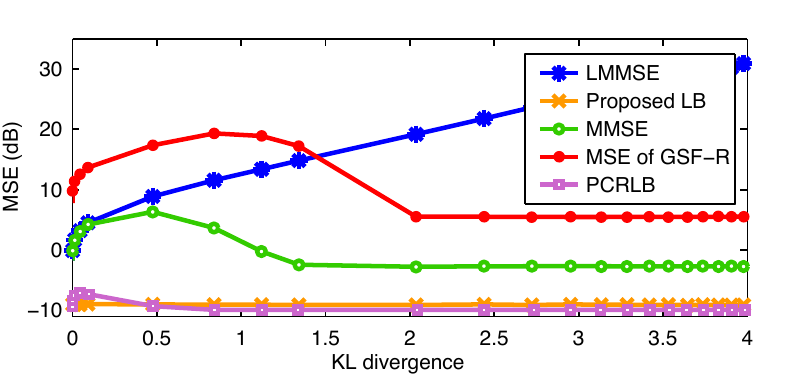}
\caption{MSE for the synthetic data generated using Model 2 vs. KL divergence between the measurement noise pdf and its moment-matched Gaussian.}
\label{fig:MSEM2}
\end{figure}

\begin{figure}[!t]
\centering
\includegraphics[width=3.5in]{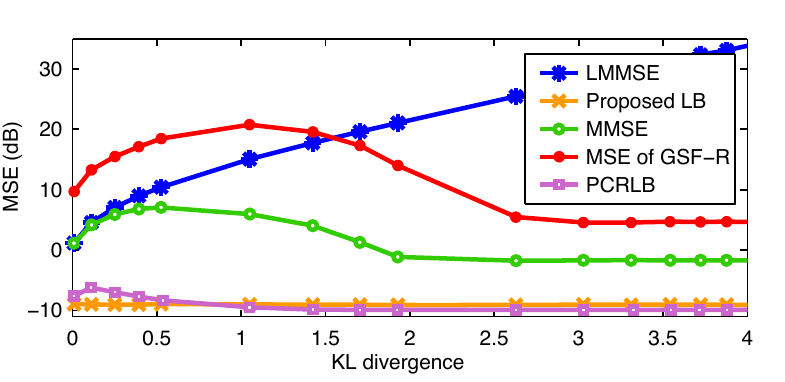}
\caption{MSE for the synthetic data generated using Model 3 vs. KL divergence between the measurement noise pdf and its moment-matched Gaussian.}
\label{fig:MSEM3}
\end{figure}

Fig.~\ref{fig:MSEM1}--\ref{fig:MSEM3}, show the MMSE versus the Kullback-Leibler (KL) divergence between the GM measurement noise distribution and its moment-matched Gaussian pdf for the three GM models used.\footnote{The KL divergence between the process noise distribution and its moment-matched Gaussian pdf for Model 1, 2, and 3 is \(2.0352\), \(1.9980\), and \(2.6255\), respectively.} \corcol{In addition to the proposed bounds we have also provided numerically approximated Posterior Cramer-Rao lower bound for the MMSE filter, for comparison purposes.} The values in these figures are evaluated at approximately \(\%95\) confidence interval with 1000 Monte Carlo runs. As shown in Section~\ref{sec:LB}, the lower bound for the MMSE is the MSE of the Matched filter, for which we have prior information about the noise clusters in effect. Since, the unconditional MSE of Matched filter is only a function of the coefficients and the covariance matrices of the clusters (see~\eqref{eq:LBMat}), it remains constant throughout our simulations, as we are only varying the separation between the clusters by changing the parameter \(c\). \corcol{This is not true for the Posterior Cramer-Rao lower bound (PCRLB). Specifically, PCRLB follows the trend of the MMSE more closely when compared with our proposed lower bound. Consequently, for smaller values of KL divergence (approximately \(\mathrm{KL}<1\)), it provides a tighter lower bound on the MMSE compared with our proposed lower bound. However, for larger values of KL divergence (approximately \(\mathrm{KL}>1\)), PCRLB is smaller than the proposed lower bound. Moreover, it is worth noting that unlike the proposed lower bound, PCRLB cannot be evaluated analytically for GSF and requires numerical approximations.} 

Unlike the proposed lower bound, the performance of the upper bounds depends on the separation between the clusters. Specifically, for smaller values of KL divergence, i.e. \(\mathrm{KL}<1.2\), \(\mathrm{KL}<1.5\), \(\mathrm{KL}<1.6\) for Model 1, 2, and 3, respectively, \corcol{the }Kalman filter provides a tighter upper bound (LMMSE) when compared with GSF-R. However, further increasing the separation between the clusters, hence the KL divergence, improves the performance of GSF-R as an upper bound. This is because with more separation between the clusters, the parameter \(\gamma_k^{ij,lm*}{\left(\mathbf{z}_{k}\right)}\) will have a smaller value for \(ij \neq lm*\). Consequently, the part of \(P_{k|k}^{\mathrm{R}}\) which depends on the distance between clusters will be smaller.


It is also worth noting that the MSE of GSF-R follows the same trend as the MMSE versus the KL divergence between the measurement noise and its moment-matched Gaussian pdf. This is due to the fact that the performance of both filters depends on the separation among the clusters, since they both depend on the parameters \(\mu_k^{ij}{\left(\mathbf{z}_{k}\right)}\) to estimate the state. Specifically, when the clusters are very close (approximately \(\mathrm{KL}<0.1\)), both MMSE and GSF-R cannot find the active models accurately as the likelihoods for all clusters are very close. Moreover, since the distribution is very close to a single Gaussian, \corcol{the }Kalman filter performs similar to the MMSE filter. However, further increasing the KL divergence, increases the MMSE until it reaches its maximum, at \corcol{\( 0.5 \leq \mathrm{KL} \leq 1\)}. This is because at lower separation levels, i.e. \(0.1 < \mathrm{KL} < 0.5\), the MMSE filter cannot differentiate between the active model and the other models using the coefficients. Increasing the KL divergence past the peak of MMSE (\(\mathrm{KL}> 0.5\)), results in lower MMSE since with higher separation, the coefficients are better indicative of the active model. This is why the MSE of GSF-R also decreases at these KL divergence values.

As mentioned earlier, to avoid having an unbounded error, in this work we used a constant process noise pdf. However, as shown in~\cite{pishdad_approximate_2014}, at very large KL divergences, e.g. \(\mathrm{KL}> 6\) for process and measurement noise, the MMSE converges to the MSE of Matched filter, i.e. the lower bound (see Proposition~\ref{prop:ConvergenceofBounds}). However, since the wrong selection of the active model can lead to unbounded error~\cite{bar2001estimation}, GSF-R becomes highly unstable for large values of KL divergence, especially for Model 1 and 3 which have a more uniform weight distribution (among the clusters of the noise distributions), and rely more on the likelihood for the correct selection of the active model.

Based on the simulation results provided in this section, the proposed analytic bounds for MMSE show good performance, with the maximum distance between the bounds and MMSE equal to 20 dB for lower bound and 10 dB for the upper bounds. Moreover, depending on the GM noise distributions in the system under study, a tighter upper bound can be selected. \corcol{For instance if the fitted Gaussians provide better representations of the noise distributions than their components with  maximum weights, LMMSE can provide a tighter bound. This is true at approximately \(KL \leq 1.5\) in our simulations.} This is particularly important since the proposed upper bounds are the MSEs of actual implementable filters. Hence, when the upper bound for the MMSE provides the desired estimation precision, the suitable upper bound filter (Kalman or GSF-R depending on the noise models) can be used instead of the MMSE filter to save computational resources. 

\section{Conclusion}
\label{sec:Conclusion}
Positioning problems can be represented using state-space systems, with the state of the system denoting the unknown location. Using this representation, Bayesian tracking techniques can be applied to the available noisy sensor measurements to estimate the location. For linear dynamic systems with Gaussian noise, \corcol{the }Kalman filter provides the MMSE state estimation. Thus, by approximating non-Gaussian  noise distributions with Gaussian Mixtures (GM), a bank of Kalman filters or Gaussian Sum Filter (GSF) can be used to find the MMSE state estimation. However, the MMSE itself is not analytically tractable. Specifically, the proposed analytic bounds in the literature, including Posterior Cramer-Rao~\cite{tichavsky_posterior_1998}, Bobrovsky-Zakai~\cite{bobrovsky_lower_1976}, and Weiss-Weinstein~\cite{weiss_lower_1985,xaver_analytic_2013}, do not have a closed-form for GM noise distributions. Hence in this work we first evaluate the MSE of GSF, and then propose analytically tractable lower and upper bounds for the MMSE.

The proposed lower bound is the MSE of a filter which only selects and uses the Kalman filter in GSF corresponding to the \textit{active model}, by using prior information. The upper bounds, however, are given by filters that do not require prior information and therefore they are implementable: We propose two filters with their MSEs acting as upper bounds for the MMSE. The first upper bound is the MSE of \corcol{the }Kalman filter which is the Linear MMSE (LMMSE). The second upper bound, is the MSE of GSF-R which only selects and uses the Kalman filter with the maximum weight in GSF. The choice between these two filters depends on the GM noise parameters, as shown through simulations. We also show that in the limit, when the Mahalanobis distances between the clusters of the posterior approach infinity, the MSE of GSF-R (upper bound) and the MMSE both converge to the lower bound. The fact that the upper bound filters are implementable is particularly important, since for  certain applications where the upper bound provides the desired accuracy, the suitable upper bound filter can be used instead of the MMSE filter, hence saving computational resources.

\appendices


\section{Proof of Lemma 1}
\label{sec:uncondMSEProof}
\begin{IEEEproof} Assume a sequential state estimator, which provides an estimate of the unknown state \({\mathbf{{x}}}_{k}\), using the available measurements \({\mathbf{{z}}}_{1:k}\). If we denote the state estimation by \(\hat{\mathbf{{x}}}_{k}\), and the state estimation error covariance matrix by \(P_{k|k}\), the unconditional MSE of this estimator, \(\epsilon_k ^2\), can be evaluated as
\corcol{
\begin{align}
\epsilon_k ^2 = & \mathds{E}_{\mathbf{x}_k,\mathbf{z}_{1:k}}{\left\lbrace \left(\mathbf{x}_{k}-\hat{\mathbf{{x}}}_{k}\right)\tran\left(\mathbf{x}_{k}-\hat{\mathbf{x}}_{k}\right) \right\rbrace }
\\= & \expectz {\left\lbrace \expectx {\left\lbrace \left(\mathbf{{x}}_{k} -  \hat{\mathbf{{x}}}_{k}\right)\tran\left(\mathbf{{x}}_{k} -  \hat{\mathbf{{x}}}_{k}\right) \big|\mathbf{{z}}_{1:k}\right\rbrace} \right\rbrace}\\
\nonumber
= & \int \int \left(\mathbf{{x}}_{k} -  \hat{\mathbf{{x}}}_{k}\right)\tran\left(\mathbf{{x}}_{k} -  \hat{\mathbf{{x}}}_{k}\right)  \\
& \times  p{\left(\mathbf{{x}}_{k}|\mathbf{{z}}_{1:k}\right)}p{\left(\mathbf{{z}}_{1:k}\right)}\mathrm{d}\mathbf{{x}}_{k}\mathrm{d}\mathbf{{z}}_{1:k} \\
= & \int \mathrm{tr}{\left(P_{k|k}\right)}p{\left(\mathbf{{z}}_{k}|\mathbf{{z}}_{1:k-1}\right)} \mathrm{d}\mathbf{{z}}_{k}. \ \IEEEQEDhere
\end{align}}
\end{IEEEproof}


\section{Evaluation of \(\mathcal{M}^{\mathrm{UB}}_k\)}
\label{sec:MUB}
In this section we provide the details on the evaluation of \(\mathcal{M}^{\mathrm{R}}_k\).

\corcol{Using~\eqref{eq:MUB} and ~\eqref{eq:defCGSFR}, we have
\begin{align}
\mathcal{M}^{\mathrm{R}}_k=& \int { \bigg( \Pmmse + \mathsf{\hat{\boldsymbol{x}}}_{k|k}\mathsf{\hat{\boldsymbol{x}}}_{k|k}\tran \bigg.}\\
&- \mathsf{\hat{\boldsymbol{x}}}_{k|k} \sum\limits_{i,j} \mathds{1}_{\mathcal{R}^{ij}_k}{\left(\mathbf{z}_k\right)} \mathsf{\hat{\boldsymbol{x}}}{_{k|k}^{ij}}\tran \\
&- \sum\limits_{i,j} \mathds{1}_{\mathcal{R}^{ij}_k}{\left(\mathbf{z}_k\right)} \mathsf{\hat{\boldsymbol{x}}}{_{k|k}^{ij}} \mathsf{\hat{\boldsymbol{x}}}_{k|k} \tran \\
& \bigg.+\sum\limits_{i,j}  \mathds{1}_{\mathcal{R}^{ij}_k} \mathsf{\hat{\boldsymbol{x}}}{_{k|k}^{ij}}\mathsf{\hat{\boldsymbol{x}}}{_{k|k}^{ij}}\tran \bigg)p{\left(\mathbf{{z}}_{k}|\mathbf{{z}}_{1:k-1}\right)}\mathrm{d}\mathbf{{z}}_{k}.
\end{align}
Hence, defining
\begin{align}
\mathcal{M}^{\mathrm{R}^1}_k \triangleq & \int{\mathsf{\hat{\boldsymbol{x}}}_{k|k} \sum\limits_{i,j} \mathds{1}_{\mathcal{R}^{ij}_k}{\left(\mathbf{z}_k\right)} \mathsf{\hat{\boldsymbol{x}}}{_{k|k}^{ij}}\tran}p{\left(\mathbf{{z}}_{k}|\mathbf{{z}}_{1:k-1}\right)}\mathrm{d}\mathbf{{z}}_{k}, \\
\mathcal{M}^{\mathrm{R}^2}_k \triangleq & \int{ \sum\limits_{i,j} \mathds{1}_{\mathcal{R}^{ij}_k}{\left(\mathbf{z}_k\right)} \mathsf{\hat{\boldsymbol{x}}}{_{k|k}^{ij}} \mathsf{\hat{\boldsymbol{x}}}_{k|k} \tran}p{\left(\mathbf{{z}}_{k}|\mathbf{{z}}_{1:k-1}\right)}\mathrm{d}\mathbf{{z}}_{k},\\
\mathcal{M}^{\mathrm{R}^3}_k \triangleq & \int{\sum\limits_{i,j}  \mathds{1}_{\mathcal{R}^{ij}_k} \mathsf{\hat{\boldsymbol{x}}}{_{k|k}^{ij}}\mathsf{\hat{\boldsymbol{x}}}{_{k|k}^{ij}}\tran}p{\left(\mathbf{{z}}_{k}|\mathbf{{z}}_{1:k-1}\right)}\mathrm{d}\mathbf{{z}}_{k},
\end{align}
and using~\eqref{eq:defM},~\eqref{eq:decomposeM}, and~\eqref{eq:defM3} we can write
\begin{align}
\mathcal{M}^{\mathrm{R}}_k=& \mathcal{M}_k^1+\mathcal{M}_k^2 -\mathcal{M}^{\mathrm{R}^1}_k-\mathcal{M}^{\mathrm{R}^2}_k +\mathcal{M}^{\mathrm{R}^3}_k.
\end{align}
To evaluate \(\mathcal{M}^{\mathrm{R}^1}_k\), we can write:
\begin{align}
\nonumber
\mathcal{M}^{\mathrm{R}^1}_k = & \sum\limits_{i,j} \sum\limits_{l,m} \int \mathds{1}_{\mathcal{R}^{ij}_k}{\left(\mathbf{z}_k\right)}\mu_k^{lm}{\left(\mathbf{z}_{k}\right)} \mathsf{\hat{\boldsymbol{x}}}_{k|k}^{lm}\mathsf{\hat{\boldsymbol{x}}}{_{k|k}^{ij}}\tran \\ & \times p{\left(\mathbf{{z}}_{k}|\mathbf{{z}}_{1:k-1}\right)}\mathrm{d}\mathbf{{z}}_{k} \\
\nonumber
=& \sum\limits_{i,j} \sum\limits_{l,m}  \mathcal{W}^l_k\mathcal{P}^m_k \int \mathds{1}_{\mathcal{R}^{ij}_k}{\left(\mathbf{z}_k\right)}\mathcal{N}\left(\mathbf{z}_k; \mathbf{\hat{z}}_{k}^{lm}, S_k^{lm}\right)\\ & \times \mathsf{\hat{\boldsymbol{x}}}_{k|k}^{lm}\mathsf{\hat{\boldsymbol{x}}}{_{k|k}^{ij}}\tran \mathrm{d}\mathbf{{z}}_{k} \\
= & \sum\limits_{i,j} \sum\limits_{l,m}  \mathcal{W}^l_k\mathcal{P}^m_k \left(\mathbf{\hat{{x}}}_{k|k-1}^{l}\mathbf{\hat{{x}}}{_{k|k-1}^{i}}\tran\right.\\
&+ \mathbf{\hat{{x}}}_{k|k-1}^{l}\left( \mathbf{\tilde{z}}_{k}^{lm} - \mathbf{\hat{z}}_{k}^{ij} \right)\tran \mathsf{W}{_k^{ij}}\tran \\
&+ \mathsf{W}_k^{lm}\left( \mathbf{\tilde{z}}_{k}^{lm} - \mathbf{\hat{z}}_{k}^{lm} \right)\mathbf{\hat{{x}}}{_{k|k-1}^{i}}\tran \\
&\left.+ \mathsf{W}_k^{lm}\left(\tilde{S}_k^{lm} + \left(\mathbf{\tilde{z}}_{k}^{lm} - \mathbf{\hat{z}}_{k}^{lm}\right) \left(\mathbf{\tilde{z}}_{k}^{lm} - \mathbf{\hat{z}}_{k}^{ij}\right) \tran \right)\mathsf{W}{_k^{ij}}\tran \right),
\end{align}
where
\begin{align}
\mathbf{\tilde{z}}_{k}^{lm} \triangleq & \int  \mathds{1}_{\mathcal{R}^{ij}_k}{\left(\mathbf{z}_k\right)}\mathbf{{z}}_{k}\mathcal{N}\left(\mathbf{z}_k; \mathbf{\hat{z}}_{k}^{lm}, S_k^{lm}\right) \mathrm{d}\mathbf{{z}}_{k} \\
\tilde{S}_k^{lm} \triangleq & \int  \mathds{1}_{\mathcal{R}^{ij}_k}{\left(\mathbf{z}_k\right)}\left(\mathbf{\mathbf{z}_k-\tilde{z}}_{k}^{lm}\right) \left(\mathbf{\mathbf{z}_k-\tilde{z}}_{k}^{lm}\right) \tran \\
& \times \mathcal{N}\left(\mathbf{z}_k; \mathbf{\hat{z}}_{k}^{lm}, S_k^{lm}\right) \mathrm{d}\mathbf{{z}}_{k}.
\end{align}
The integrals \(\mathcal{M}^{\mathrm{R}^2}_k\) and \(\mathcal{M}^{\mathrm{R}^3}_k\) can be evaluated in a similar manner.}

%


\section*{Acknowledgment}

This work was partly supported by the Natural Sciences and Engineering Research Council (NSERC) and industrial and government partners, through the Healthcare Support through Information Technology Enhancements (hSITE) Strategic Research Network.

\ifCLASSOPTIONcaptionsoff
  \newpage
\fi



%

\bibliographystyle{IEEEtran}
\bibliography{KFBank}

%

%






\end{document}